%% file: epi.tex
\documentstyle[12pt,epsfig]{article}

\expandafter\ifx\csname mathrm\endcsname\relax\def\mathrm#1{{\rm #1}}\fi

\makeatletter
\if@twoside \oddsidemargin -17pt \evensidemargin 00pt
\else \oddsidemargin 00pt \evensidemargin 00pt
\fi
\expandafter\ifx\csname mathrm\endcsname\relax\def\mathrm#1{{\rm #1}}\fi

\makeatother

\def\beq{\begin{equation}}
\def\eeq{\end{equation}}
\def\beqar{\begin{eqnarray}}
\def\eeqar{\end{eqnarray}}
\def\barr#1{\begin{array}{#1}}
\def\earr{\end{array}}
\def\bfi{\begin{figure}}
\def\efi{\end{figure}}
\def\btab{\begin{table}}
\def\etab{\end{table}}
\def\bce{\begin{center}}
\def\ece{\end{center}}
\def\nn{\nonumber}

\def\text{\textstyle}

\def\al{\alpha}

\def\de{\delta}

\def\De{\Delta}

\def\refeq#1{\mbox{(\ref{#1})}}
\def\reffi#1{\mbox{Fig.~\ref{#1}}}

\def\citere#1{\mbox{Ref.~\cite{#1}}}
\def\citeres#1{\mbox{Refs.~\cite{#1}}}

\newcommand{\GeV}{\unskip\,{\mathrm GeV}}
\newcommand{\MeV}{\unskip\,{\mathrm MeV}}
\newcommand{\TeV}{\unskip\,{\mathrm TeV}}

\def\mathswitchr#1{\relax\ifmmode{\mathrm{#1}}\else$\mathrm{#1}$\fi}

\newcommand{\PW}{\mathswitchr W}
\newcommand{\PZ}{\mathswitchr Z}

\newcommand{\PH}{\mathswitchr H}

\newcommand{\Pb}{\mathswitchr b}

\newcommand{\Pt}{\mathswitchr t}

\def\mathswitch#1{\relax\ifmmode#1\else$#1$\fi}

\newcommand{\MW}{\mathswitch {M_\PW}}

\newcommand{\MZ}{\mathswitch {M_\PZ}}
\newcommand{\MH}{\mathswitch {M_\PH}}

\newcommand{\Mb}{\mathswitch {m_\Pb}}
\newcommand{\Mt}{\mathswitch {m_\Pt}}

\newcommand{\scrs}{\scriptscriptstyle}
\newcommand{\sw}{\mathswitch {s_{\scrs\PW}}}
\newcommand{\swtwo}{\mathswitch {s_{{\scrs\PW}, (2)}}}
\newcommand{\swtone}{\mathswitch {s_{{\scrs\PW}, (1), \Pt}}}
\newcommand{\swHone}{\mathswitch {s_{{\scrs\PW}, (1), \PH}}}

\newcommand{\GF}{\mathswitch {G_\mu}}

\newcommand{\alps}{\alpha_{\mathrm s}}

\newcommand{\ses}{self-en\-er\-gies}
\newcommand{\fea}{{\em FeynArts}}

\newcommand{\two}{{\em TwoCalc}}

\renewcommand{\Re}{\mathop{\mathrm{Re}}}

\begin{document}
\thispagestyle{empty}
\null
\hfill KA-TP-06-1997\\
\null
\hfill hep-ph/9708321\\
\vskip .8cm
\begin{center}
{\Large \bf Precise Predictions for the W-Boson Mass%
\footnote{Presented at the Cracow Epiphany Conference on W Boson,
Cracow, January 4--6, 1997.}
}
\vskip 2.5em
{\large
{\sc Georg Weiglein}\\[1ex]
{\normalsize \it Institut f\"ur Theoretische Physik, Universit\"at
Karlsruhe,\\
D--76128 Karlsruhe, Germany}
}
\vskip 2em
\end{center} \par
\vskip 1.2cm
\vfil
{\bf Abstract} \par
Recent results for higher-order corrections 
to the relation between the vector-boson masses
in the Standard Model and Supersymmetry 
are summarized.
In the Standard Model, the Higgs-mass dependence of the two-loop 
contributions to $\Delta r$ is studied. 
Exact results are given for the Higgs-dependent two-loop corrections
associated with the fermions, i.e.\ no expansion in the top-quark and
the Higgs-boson mass is made. The results for the top quark are
compared with results of an expansion up to next-to-leading order in
the top-quark mass. Agreement is found within $30 \%$ of the two-loop
result.
In Supersymmetry, the two-loop QCD corrections to 
the stop- and sbottom-loop contributions to the $\rho$~parameter are 
presented. The two-loop corrections modify the one--loop contribution by
up to 30\%; the gluino decouples for large masses. Contrary to the SM
case where the QCD corrections are negative and screen the one-loop
value, the corresponding corrections in the supersymmetric case are in
general positive, increasing the sensitivity in the search for scalar
quarks through their virtual effects in high-precision electroweak
observables.
\par
\vskip 1cm
\null
\setcounter{page}{0}
\clearpage

\section{Introduction}

With the prospect of the improving accuracy of the measurement of the 
W-boson mass at LEP2 and the Tevatron, the importance
of the basic relation between the masses $\MW$, $\MZ$ of the vector
bosons, the Fermi constant $\GF$ and the fine structure constant $\al$
for testing the Standard Model (SM) and extensions of it, most
prominently the Minimal Supersymmetric Standard Model (MSSM),
becomes even more pronounced. This relation is commonly expressed in
terms of the quantity $\De r$~\cite{sirlin} derived from muon decay.
After the discovery of the top
quark~\cite{mtexp}, whose mass had already successfully been predicted 
by confronting the electroweak theory with the precision data,
an important goal for the future is to further constrain the mass
of the Higgs boson, $\MH$, for which at the moment only rather mild
bounds exist (see e.g.\ \citere{datasum96}). In order to improve on this
situation, and also to achieve a higher sensitivity to effects of
physics beyond the SM, a further reduction of the experimental and
theoretical errors is necessary.

Concerning the reduction of the theoretical error due to missing 
higher-order corrections, 
in particular a precise prediction for $\Delta r$ is
of interest. At the one-loop level the largest contributions to
$\Delta r$ in the SM are the QED induced shift in the fine structure
constant, $\De \al$, and the contribution of the top/bottom weak isospin
doublet, which gives rise to a term that grows as $\Mt^2$.
This contribution enters $\Delta r$ via the $\rho$
parameter~\cite{R7}, which measures the relative strength of the neutral
to charged current processes at zero momentum-transfer.
The SM one-loop result for $\De r$~\cite{sirlin} has been supplemented by
resummations of certain one-loop contributions~\cite{resum,sirresum}. 
While QCD corrections at ${\cal O}(\al \alps)$~\cite{R7a,qcd2} and 
${\cal O}(\al \alps^2)$~\cite{qcd3} are
available, the electroweak results at the two-loop level have so far
been restricted to expansions in either $\Mt$ or $\MH$. The leading top-quark
and Higgs-boson contributions were evaluated in
\citeres{vdBH,vdBV}. The full Higgs-boson dependence of
the leading $\GF^2 \Mt^4$ contribution was calculated in 
\citere{barb2}, and recently also
the next-to-leading top-quark contributions of 
${\cal O}(\GF^2 \Mt^2 \MZ^2)$ were derived~\cite{gamb}.

In the global SM fits to all available data (see e.g.\ \citere{datasum96}), 
where the ${\cal O}(\GF^2 \Mt^2 \MZ^2)$ correction obtained in \citere{gamb} 
is not yet included, the error due to missing higher-order corrections 
has a strong effect on the resulting value of $\MH$, shifting the upper
bound for $\MH$ at 95\% C.L.\ by $\sim +100 \GeV$. In \citeres{DGS}
it is argued that inclusion of the ${\cal O}(\GF^2 \Mt^2 \MZ^2)$ will lead
to a significant reduction of this error. 

Since both the Higgs-mass dependence of the leading $\Mt^4$ contribution
and the inclusion of the next-to-leading term in the $\Mt$~expansion
turned out to yield important corrections, in order to further settle
the issue of theoretical uncertainty due to missing higher-order
corrections a more complete calculation 
would be desirable, where no expansion in $\Mt$ or $\MH$ is made.

In the MSSM, the one-loop result for $\De r$ is
known~\cite{R6}. The most important supersymmetric (SUSY) contribution 
is that of the stop and sbottom loops to the 
$\rho$~parameter~\cite{R6rho}.
If there is a large splitting between the masses of these particles, 
in analogy to the SM case the contribution will grow with the 
squared mass of 
the heaviest scalar quark and can be sizable.
In order to treat the SUSY loop contributions to the electroweak
observables at the same level of accuracy as the standard contribution,
higher-order corrections should be incorporated. In particular the QCD 
corrections, which because of the large value of the strong coupling 
constant can be rather important, are of interest.

In this article recent results obtained in the SM and the MSSM at 
the two-loop level are summarized. In the SM, the $\MH$-dependence
of the two-loop contributions to $\De r$ is studied and
the corrections associated with the fermions are
evaluated exactly~\cite{sbaugw1,sbaugw2},
i.e.\ without an expansion in the masses. In the MSSM, results
for the two-loop QCD corrections to the $\rho$~parameter are 
presented~\cite{susydelrho}.

\section{Higgs-mass dependence of two-loop corrections to $\De r$}
\label{sect:DeltaR}

The correlation between the vector-boson masses in terms of the Fermi
constant reads~\cite{sirlin}
\beq
\MW^2 \left(1 - \frac{\MW^2}{\MZ^2}\right) = 
\frac{\pi \al}{\sqrt{2} \GF} \left(1 + \De r\right),
\eeq
where the radiative corrections are contained in the quantity $\De r$.
In the context of this paper we treat $\De r$ without resummations,
i.e.\ as being fully expanded up to two-loop order,
\beq
\De r = \De r_{(1)} + \De r_{(2)} + {\cal O}(\al^3) .
\eeq
The theoretical predictions for $\De r$ are obtained by calculating
radiative corrections to muon decay. 

{}From a technical point of view the calculation of top-quark
and Higgs-boson
contributions to $\Delta r$ and other processes with light external
fermions at low energies requires in particular the evaluation of
two-loop self-energies on-shell, 
i.e.\ at non-zero external momentum, while vertex and
box contributions can mostly be reduced to vacuum integrals. The
problems encountered in such a calculation are due to the large number
of contributing Feynman diagrams, their complicated tensor structure, 
the fact that scalar two-loop integrals are in general not expressible
in terms of polylogarithmic functions~\cite{ScharfDipl}, and due to the
need for a two-loop renormalization, which has not yet been worked out
in full detail. 

The methods that we use for carrying out such a calculation have been
outlined in \citere{sbaugw1}. The generation of the diagrams and
counterterm contributions is done with the help of the computer-algebra
program \fea\ \cite{fea}. Making use of two-loop tensor-integral 
decompositions, the generated amplitudes are reduced to a minimal set 
of standard scalar integrals with the program \two~\cite{two}. The
renormalization is performed within the complete on-shell scheme
(see e.g.\ \citere{Dehab}), i.e.\
physical parameters are used throughout. The two-loop scalar integrals
are evaluated numerically with one-dimensional integral
representations~\cite{intnum}. These allow a very fast calculation of the
integrals with high precision without any approximation in the masses.

As an application, we study here the Higgs-mass dependence of different
two-loop contributions to $\De r$. To this end we consider the
subtracted quantity
\beq
\label{eq:DeltaRsubtr}
\De r_{(2), {\mathrm subtr}}(\MH) =
\De r_{(2)}(\MH) - \De r_{(2)}(\MH = 65\GeV),
\eeq
where $\De r_{(2)}(\MH)$ denotes the two-loop contribution to
$\De r$.

\subsection{Higgs-mass dependence of two-loop top-quark contributions}

Potentially large $\MH$-dependent contributions are the corrections
associated with the top quark, since the Yukawa coupling of the Higgs 
to the top quark is proportional to $\Mt$, and the contributions which
are proportional to $\De\al$. We first consider the
Higgs-mass dependence of the two-loop top-quark contributions
and calculate the quantity $\De r^{\mathrm top}_{(2), {\mathrm
subtr}}(\MH)$ which denotes the contribution of the top/bottom doublet
to $\De r_{(2), {\mathrm subtr}}(\MH)$.

{}From the one-particle irreducible diagrams obviously 
those graphs contribute
to $\De r^{\mathrm top}_{(2), {\mathrm subtr}}(\MH)$ that contain both
the top quark and the Higgs boson. It is easy to see that only two-point
functions enter in this case, since all graphs where the Higgs boson
couples to the muon or the electron may safely be neglected.
Although no two-loop three-point function enters, there is
nevertheless a contribution from the two-loop and one-loop vertex
counterterms. If the field renormalization constants of the W~boson
are included (which cancel in the complete result), the vertex
counterterms are separately finite.

Expressed in terms of the one-loop and two-loop contributions to 
the transverse part of the W-boson self-energy 
$\Sigma^{\PW}(p^2)$ and the counterterm
$\de Z^{\mathrm vert}$ to the $W^- \bar e \nu_e$ vertex the quantity 
$\De r^{\mathrm top}_{(2), {\mathrm subtr}}(\MH)$ reads
\beqar
\lefteqn{
\De r^{\mathrm top}_{(2), {\mathrm subtr}}(\MH) = 
\biggl[ \frac{\Sigma^{\PW}_{(2)}(0) - 
\Re \Sigma^{\PW}_{(2)}(\MW^2)}{\MW^2} 
+ 2 \de Z^{\mathrm vert}_{(2)}  \nn } \\
&& {} + 2 \frac{\left(\Sigma^{\PW}_{(1), \Pt}(0) -
\Re \Sigma^{\PW}_{(1), \Pt}(\MW^2)\right) 
\left(\Sigma^{\PW}_{(1), \PH}(0) -
\Re \Sigma^{\PW}_{(1), \PH}(\MW^2)\right)}{\MW^4} \nn \\
&& {} + 2 \frac{\left(\Sigma^{\PW}_{(1), \Pt}(0) - 
\Re \Sigma^{\PW}_{(1), \Pt}(\MW^2)\right) 
\de Z^{\mathrm vert}_{(1), \PH}}{\MW^2}  \nn \\
&& {} + 2 \frac{\left(\Sigma^{\PW}_{(1), \PH}(0) - 
\Re \Sigma^{\PW}_{(1), \PH}(\MW^2)\right)
\de Z^{\mathrm vert}_{(1), \Pt}}{\MW^2} 
+ 2 \de Z^{\mathrm vert}_{(1), \Pt}
\de Z^{\mathrm vert}_{(1), \PH}
\biggr]_{\mathrm subtr}, 
\label{eq:Deltrtopsubtr}
\eeqar
where it is understood that the two-loop contributions to the \ses\
contain the subloop renormalization. The two-loop terms denote those
graphs that contain both the top quark and the Higgs boson, while for the 
one-loop terms the top-quark and the Higgs-boson contributions are
indicated by a subscript. The two-loop vertex counterterm is expressible
in terms of the charge counterterm $\de Z_e$ and the mixing-angle
counterterm $\de \sw/\sw$,
\beq
\de Z^{\mathrm vert}_{(2)} = \de Z_{e, (2)} - \frac {\de \swtwo}{\sw}
+ 2 \frac {\de \swtone}{\sw} \frac {\de \swHone}{\sw} 
- \de Z_{e, (1), \Pt} \frac {\de \swHone}{\sw},
\label{eq:dZvert2}
\eeq
and analogously for the one-loop vertex counterterm.
For the considered contributions the charge counterterm is related to
the photon vacuum polarization according to~\cite{bfmlong}
\beq
\de Z_{e, (2)} = - \frac{1}{2} \de Z_{AA, (2)} = 
\frac{1}{2} \Pi^{AA}_{(2)}(0),
\label{eq:dZe2}
\eeq
and similarly to the one-loop case
the mixing angle counterterm $\de \swtwo/\sw$ is expressible 
in terms of the on-shell two-loop W-boson and Z-boson \ses\ and additional
one-loop contributions~\cite{sbaugw2}. In \refeq{eq:Deltrtopsubtr} and 
\refeq{eq:dZvert2}
the field renormalization constants of the W~boson have been omitted.
In our calculation of $\De r^{\mathrm top}_{(2), {\mathrm subtr}}(\MH)$
we have explicitly kept the field renormalization constants of
all internal fields and have checked
that they actually cancel in the final result.

\begin{figure}[htb]
\begin{center}
\input{DELTAR2}
\end{center}
\caption{
Two-loop top-quark contribution to
$\Delta r$ subtracted at $\MH=65\,\GeV$.}
\label{fig:delr1}
\end{figure}
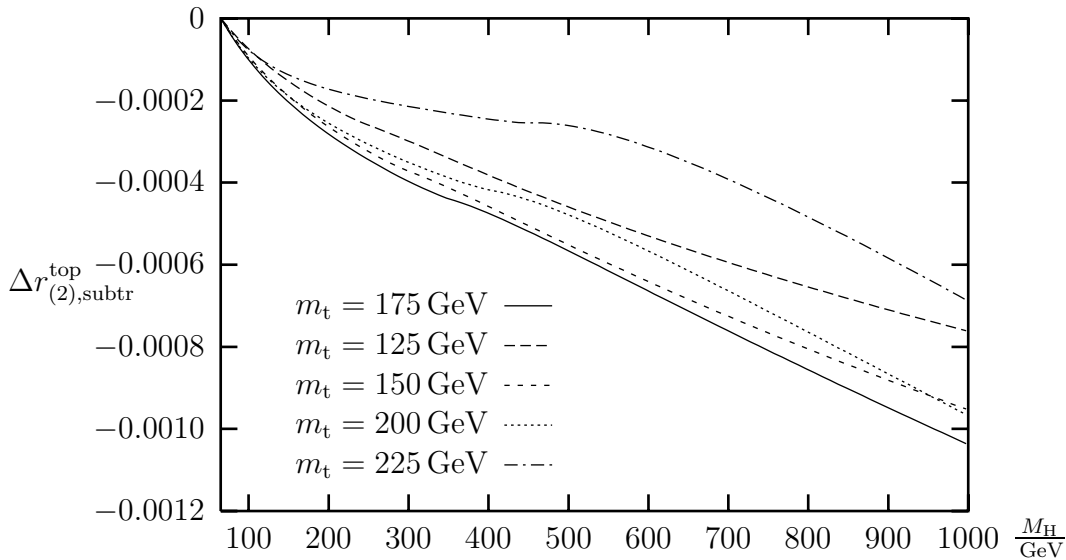

The result for $\De r^{\mathrm top}_{(2), {\mathrm subtr}}(\MH)$
is shown in \reffi{fig:delr1} for various values of $\Mt$. The
Higgs-boson mass is varied in the interval 
$65 \GeV \leq \MH \leq 1 \TeV$. The change in 
$\De r^{\mathrm top}_{(2), {\mathrm subtr}}(\MH)$ over this interval
is about 0.001, which corresponds to a shift in $\MW$ of about 
$20 \MeV$.
It is interesting to note that the absolute value of
the correction is maximal just in the region of $\Mt = 175 \GeV$,
i.e.\ for the physical value of the top-quark mass. 
For $\Mt \sim 175 \GeV$ the correction 
$\De r^{\mathrm top}_{(2), {\mathrm subtr}}(\MH)$
amounts to about $10 \%$ of the one-loop contribution,
$\De r_{(1), {\mathrm subtr}}(\MH)$,
which is defined in analogy to~\refeq{eq:DeltaRsubtr}.

\pagebreak
\subsection{Higgs-mass dependence of 
the other fermionic contributions}

The other $\MH$-dependent two-loop correction that is expected to
be sizable is 
the contribution of the terms proportional to $\De \al$. It reads
\beqar
\De r^{\De\al}_{(2), {\mathrm subtr}}(\MH) & = &
2 \De\al \left[
\frac{\Sigma^{\PW}_{(1), \PH}(0) -
\Re \Sigma^{\PW}_{(1), \PH}(\MW^2)}{\MW^2} - 
2 \frac {\de \swHone}{\sw} \right]_{\mathrm subtr} \nn \\
&=& 2 \De\al \, \De r_{(1), {\mathrm subtr}}(\MH),
\label{eq:DelRbarDelAlp}
\eeqar
and can easily be obtained by a proper resummation of one-loop
terms~\cite{sirresum}. 

The remaining fermionic contribution,
$\De r^{\mathrm lf}_{(2), {\mathrm subtr}}$,
is the one of the light fermions,
i.e.\ of the leptons and of the quark doublets of the first and second
generation,
which is not contained in $\De\al$. Its structure is analogous to
\refeq{eq:Deltrtopsubtr}, but due to the negligible coupling of
the light fermions to the Higgs boson much less diagrams contribute.


\begin{figure}[htb]
\begin{center}
\input{DELTARMHDEP}
\end{center}
\caption{
One-loop and two-loop contributions to $\Delta r$
subtracted at $\MH=65\,\GeV$.
$\De r_{\mathrm subtr}$ is the result for the full one-loop and
fermionic two-loop contributions to $\De r$, as defined in the text.
\label{fig:delr2}}
\end{figure}
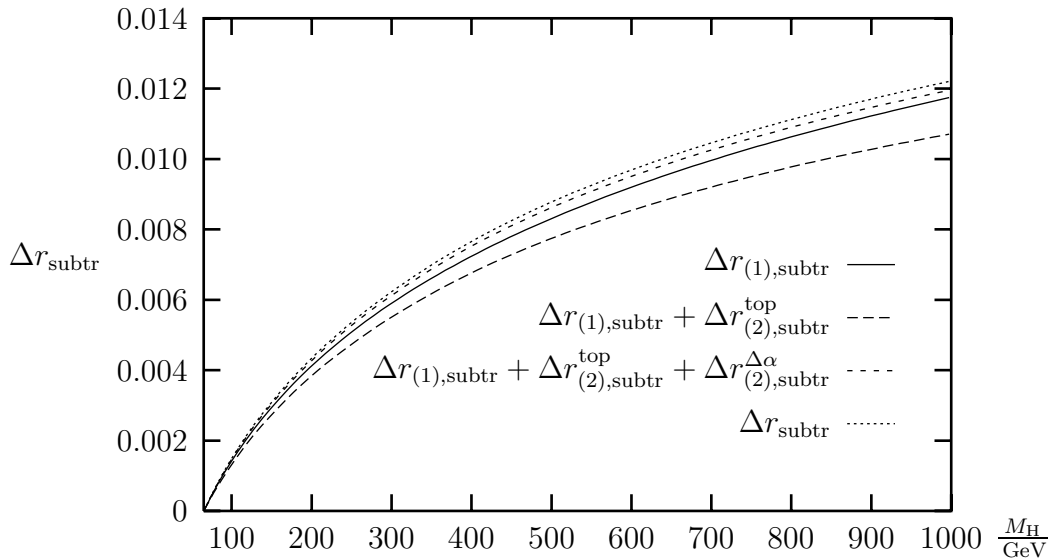

The total result for the one-loop and fermionic two-loop contributions
to $\De r$, subtracted at $\MH=65\,\GeV$, reads
\beq
\De r_{\mathrm subtr} \equiv \De r_{(1), {\mathrm subtr}} +
\De r^{\mathrm top}_{(2), {\mathrm subtr}} +
\De r^{\De\al}_{(2), {\mathrm subtr}} +
\De r^{\mathrm lf}_{(2), {\mathrm subtr}} .
\eeq
It is shown in \reffi{fig:delr2}, where separately also
the one-loop contribution
$\De r_{(1), {\mathrm subtr}}$, as well as
$\De r_{(1), {\mathrm subtr}} + \De r^{\mathrm top}_{(2), {\mathrm
subtr}}$, and
$\De r_{(1), {\mathrm subtr}} + \De r^{\mathrm top}_{(2), {\mathrm
subtr}} + \De r^{\De\al}_{(2), {\mathrm subtr}}$
are shown for $\Mt = 175.6 \GeV$.
It can be seen that the higher-order contributions
$\De r^{\mathrm top}_{(2), {\mathrm subtr}}(\MH)$ and
$\De r^{\De\al}_{(2), {\mathrm subtr}}(\MH)$ are of about the same size
and to a large extent cancel each other.
If the one-loop result had only been supplemented by the
contribution~\refeq{eq:DelRbarDelAlp}, which is accessible by
resummation of one-loop quantities, 
but not by the contribution of the 
irreducible two-loop diagrams, 
the result for the Higgs-mass dependence would have been misleading.
The light-fermion contributions which are not contained in $\De\al$
add a relatively small correction. Over the full range of the
Higgs-boson mass it amounts to about $4\, \MeV$.
In total, the inclusion of the higher-order contributions discussed here
leads to a slight increase in the sensitivity to the Higgs-boson mass
compared to the pure one-loop result.

Regarding the remaining Higgs-mass dependence of $\De r$ at the
two-loop level, there are only purely bosonic corrections left, which
contain no specific source of enhancement. They
can be expected to yield a contribution to
$\De r_{(2), {\mathrm subtr}}(\MH)$ of about the same size as
$\left.\left(\De r^{\mathrm bos}_{(1)}(\MH)\right)^2\right|_{\mathrm
subtr}$, where $\De r^{\mathrm bos}_{(1)}$ denotes the bosonic
contribution to $\De r$ at the one-loop level. The contribution of
$\left.\left(\De r^{\mathrm bos}_{(1)}(\MH)\right)^2\right|_{\mathrm
subtr}$ amounts to only about $10 \%$ of $\De r^{\mathrm top}_{(2),
{\mathrm subtr}}(\MH)$ corresponding to a shift of about $2 \MeV$ in
the W-boson mass.



\subsection{Comparison with an expansion in $\Mt$}

The result for
$\De r_{\mathrm subtr}^{{\mathrm top}, \De\al} \equiv
\De r_{(1), {\mathrm subtr}} + \De r^{\mathrm top}_{(2), {\mathrm
subtr}} + \De r^{\De\al}_{(2), {\mathrm subtr}}$
can be compared to the result
obtained via an expansion in $\Mt$ up
to next-to-leading order, i.e.\
${\cal O}(\GF^2 \Mt^2 \MZ^2)$ \cite{gamb,DGS}.
{}From this expansion the results for $\MW$ as a function of $\MH$ 
read (without QCD corrections; $\Mt = 175.6$)~\cite{gambpriv}
\beq
\label{tab:MWgamb}
\barr{|c||c|c|c|c|c|} \hline
\MH/\GeV & 65 & 100 & 300 & 600 & 1000\\ \hline
\MW/\GeV & 80.4819 & 80.4584 & 80.3837 & 80.3294 & 80.2901 \\ \hline
\earr
\quad .
\eeq

Extracting from \refeq{tab:MWgamb} the corresponding values of 
$\De r$ and subtracting at $\MH = 65\, \GeV$ yields the values
$\De r_{\mathrm subtr}^{{\mathrm top}, \De\al, {\mathrm expa}}(\MH)$
as results of the expansion in $\Mt$.
The comparison with the exact result 
$\De r_{\mathrm subtr}^{{\mathrm top}, \De\al}(\MH)$
reads
\beq
\label{tab:MWcomp}
\barr{|c||c|c||c|} \hline
\MH/\GeV &
\De r_{\mathrm subtr}^{{\mathrm top}, \De\al}/10^{-3} &
\De r^{{\mathrm top}, \De\al, {\mathrm expa}}_{\mathrm subtr}/10^{-3} &
\de \MW/\MeV\\ \hline
65   & 0    & 0    & 0   \\ \hline
100  & 1.48 & 1.52 & 0.6 \\ \hline
300  & 6.16 & 6.32 & 2.5 \\ \hline
600  & 9.56 & 9.79 & 3.6 \\ \hline
1000 & 12.0\phantom{00} & 12.3\phantom{00} & 4.1 \\ \hline
\earr
\quad ,
\eeq
where in the last column the approximate shift in $\MW$ is given that
corresponds to the difference between exact result and expansion.
The results agree within about $30 \%$ of $\De r^{\mathrm top}_{\mathrm
subtr}(\MH)$, which amounts to a difference in $\MW$ of up to about 
$4\, \MeV$.

\section{QCD corrections to the $\rho$~parameter in the MSSM}

The leading contributions to the $\rho$~parameter can be written
in terms of the transverse parts of the W- and Z-boson \ses\ at zero
momentum-transfer,
\beq
\Delta \rho = 
\frac{\Sigma^{\PZ\PZ}(0)}{\MZ^2} - \frac{\Sigma^{\PW\PW}(0)}{\MW^2}  \ .
\eeq
In the SM, the contribution of a fermion isodoublet $(u,d)$
to $\Delta \rho$ reads at one-loop order
\beq
\Delta \rho_0^{\rm SM} = \frac{N_c \GF}{8 \sqrt{2} \pi^2} F_0 
\left( m_u^2, m_d^2 \right) \ , 
\eeq
with the color factor $N_c$ and the function $F_0$ given by
\beq
F_0(x,y)= x+y - \frac{2xy} {x-y} \log \frac{x}{y} \ . 
\label{eq:lett3}
\eeq
The function $F_0\left( m_u^2, m_d^2 \right)$ 
vanishes if the $u$- and $d$-type quarks are 
degenerate in mass: $F_0(m_q^2, m_q^2)=0$; in the limit of large 
quark mass splitting it is proportional to the heavy quark mass 
squared: $F_0(m_q^2,0)=m_q^2$. Therefore, in the SM the only relevant
contribution is due to the top/bottom weak isodoublet. Because 
$\Mt \gg \Mb$, one obtains $\Delta \rho ^{\rm SM}_0 = 3 \GF\Mt^2/(8 
\sqrt{2} \pi^2)$. The two-loop QCD corrections in the SM
read~\cite{R7a}:
\beq
\Delta \rho ^{\rm SM}_1 = - \Delta \rho_0^{\rm SM} 
\frac{2}{3} \frac{\alpha_s}{\pi} (1+ \frac{\pi^2}{3} ) . 
\eeq

In SUSY theories, the scalar partners of each SM quark will induce 
additional contributions. The current eigenstates, 
$\tilde{q}_L$ and $\tilde{q}_R$, mix to give the mass eigenstates. 
The mixing angle is proportional to the quark mass and therefore is 
important only in the case of the third generation scalar quarks~\cite{R9}. 
In particular, due to the large value of $\Mt$, the mixing angle 
$\theta_{\tilde{\Pt}}$ between $\tilde{t}_L$ and $\tilde{t}_R$ can be 
very large and lead to a scalar top quark $\tilde{t}_1$ much 
lighter than the top quark and all the scalar partners of the light 
quarks~\cite{R9}. The mixing in the bottom-quark sector can be
sizable only in a small area of the SUSY parameter space. 

\begin{figure}[htb]
\begin{center}
\mbox{
\psfig{figure=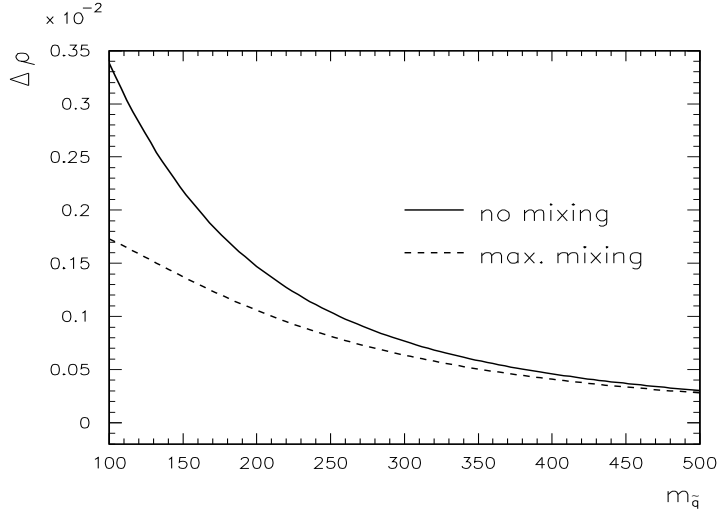,width=9cm,height=6.3cm,bbllx=140pt,bblly=285pt,bburx=450pt,bbury=535pt}}
\end{center}
\caption[]{One-loop contribution of the $(\tilde{t}, \tilde{b})$ 
doublet to $\Delta \rho$ as a function of the common mass $m_{\tilde{q}}$, 
for $\theta_{\tilde{\Pt}} =0$ and $\theta_{\tilde{\Pt}} \sim-\pi/4$
(with $\tan \beta=1.6$ and $m_{\rm LR}=0$ and 200 GeV, respectively, 
where $m_{\rm LR}$ is the off-diagonal term in the $\tilde{t}$ mass matrix).}
\label{fig:lett2}
\end{figure}

Similarly to the SM case, 
the contribution of a scalar 
quark doublet $(\tilde{u}, \tilde{d})$
vanishes if all masses are  degenerate. This means that 
in most SUSY scenarios, where the scalar partners of the light quarks 
are almost mass degenerate, only the third generation will contribute.
Neglecting the mixing in the $\tilde{b}$ sector, $\Delta \rho$ is given 
at one-loop order by the simple expression~\cite{R6rho}
\beqar
\Delta \rho ^{\rm SUSY}_0 &=& \frac{3 \GF}{8 \sqrt{2} \pi^2} \left[ -
\sin^2 \theta_{\tilde{\Pt}} \cos^2 \theta_{\tilde{\Pt}} 
F_0\left( m_{\tilde{\Pt}_1}^2,  m_{\tilde{\Pt}_2}^2 \right)
\right. \nonumber \\ 
&& \left. + \cos^2 \theta_{\tilde{\Pt}} F_0\left( m_{\tilde{\Pt}_1}^2,  
m_{\tilde{\Pb}_L}^2 \right) + \sin^2 \theta_{\tilde{\Pt}} F_0\left( 
m_{\tilde{\Pt}_2}^2,  m_{\tilde{\Pb}_L}^2 \right) \right]. 
\label{eq:lett5}
\eeqar
In a large area of the parameter space, the stop mixing angle is 
either very small, $\theta_\Pt \sim 0$, or maximal, $\theta_\Pt \sim -\pi/4$. 
The contribution $\Delta \rho_0^{\rm SUSY}$ is shown in \reffi{fig:lett2}
as a function of the common scalar mass 
$m_{\tilde{q}}=m_{\tilde{\Pt}_{L,R}} =m_{\tilde{\Pb}_{L}}$ 
(see e.g.\ \citere{DHJ})
for these two scenarios. The contribution can be at the
level of a few per mille and therefore within the range of the experimental
observability. Relaxing the assumption of a common scalar quark mass, 
the corrections can become even larger~\cite{R6rho}. 

\begin{figure}[htb]
\begin{center}
\mbox{
\psfig{figure=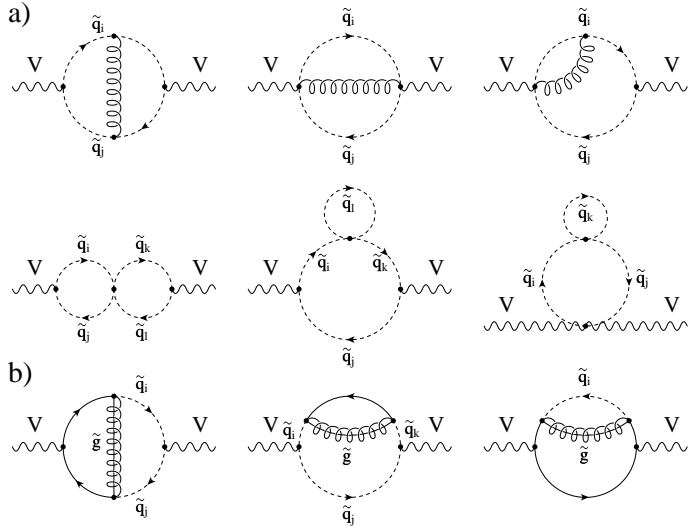,width=8.5cm,bbllx=165pt,bblly=495pt,bburx=450pt,bbury=725pt}}
\end{center}
\caption[]{Typical Feynman diagrams for the contribution of scalar 
quarks and gluinos to the W/Z-boson self-energies at the 
two-loop level.}
\label{fig:lett3}
\end{figure}

At ${\cal O}(\alpha \alpha_s)$, the two-loop Feynman diagrams contributing
to the $\rho$ parameter in the MSSM (see \reffi{fig:lett3})
consist of two sets which, at vanishing 
external momentum and after the inclusion of the counterterms, are separately 
ultraviolet finite and gauge-invariant. The first one contains 
diagrams involving only 
gluon exchange, \reffi{fig:lett3}a; in this case 
the calculation is similar to the SM, although technically more complicated 
due to the larger number of diagrams and the presence of $\tilde{q}$ mixing. 
The diagrams involving the quartic scalar-quark interaction in 
\reffi{fig:lett3}a
either contribute only to the longitudinal component of the 
self-energies or can be absorbed into the squark mass and mixing-angle 
renormalization. The renormalization of the mixing-angle
is performed in such a way that all transitions from $\tilde{q}_i 
\leftrightarrow \tilde{q}_j$ which do not depend on the loop-momenta 
in the two-loop diagrams are canceled; 
therefore the contribution of the pure scalar quark diagrams in
\reffi{fig:lett3}a is completely canceled by the renormalization.
The second set of graphs consists of 
diagrams involving scalar quarks, gluinos as well as quarks, 
\reffi{fig:lett3}b; in this case the calculation becomes 
much more complicated due to 
the even larger number of diagrams and to the presence of up to 5 particles 
with different masses in the loops. 

In order to discuss our results, let us first concentrate on the 
contribution of the gluonic corrections, 
\reffi{fig:lett3}a, and the corresponding counterterms.
At the two-loop level, the results for the electroweak gauge-boson 
self-energies at zero momentum-transfer have very simple 
analytical expressions. In the case of an isodoublet $(\tilde{u}, 
\tilde{d})$ where general mixing is allowed, the structure is similar 
to the one-loop case: 
\beqar
\Sigma^{\PW\PW} (0)
& = & - \frac{\GF \MW^2 \alpha_s}{4 \sqrt{2} \pi^3} \, \sum_{i,j=1,2} 
\left( a^{\tilde{u}}_i a^{\tilde{d}}_j \right)^2 \, F_1 \left( 
m_{\tilde{u}_i}^2, m_{\tilde{d}_j}^2 \right), \nonumber \\
\Sigma^{\PZ\PZ}(0)
& = &- \frac{\GF \MZ^2 \alpha_s}{8 \sqrt{2} \pi^3}   
\sum_{\tilde{q}= \tilde{u},\tilde{d} \atop i,j=1,2} (a_i^{\tilde{q}} 
a_j^{\tilde{q}} )^2 \, F_1 \left( m_{\tilde{q}_i}^2, m_{\tilde{q}_j}^2 
\right) ,
\end{eqnarray}
where the factors $a_{i}^{\tilde{q}}$ are given in terms of the squark
mixing angle $\theta_{\tilde{q}}$ as $a_{1}^{\tilde{q}}= \cos \theta_{
\tilde{q}}$ and $a_{2}^{\tilde{q}} = \sin \theta_{\tilde{q}}$.
The two-loop function $F_1(x,y)$ is given in terms of dilogarithms by
\beqar
F_{1}(x,y) &=& x+y- 2\frac{xy}{x-y} \log \frac{x}{y} \left[2+
\frac{x}{y} \log \frac{x}{y} \right] \nonumber \\
&& {} +\frac{(x+y)x^2}{(x-y)^2}\log^2 \frac{x}{y} 
-2(x-y) {\rm Li}_2 \left(1-\frac{x}{y} \right) . 
\eeqar
This function is symmetric in the interchange of $x$ and $y$.
As in the case of the one-loop function $F_0$, it vanishes for 
degenerate masses, $F_1(x,x)=0$, while in the case of large 
mass splitting it increases with the heavy scalar quark mass 
squared: $F_1 (x,0) = x( 1 +\pi^2/3)$.  

\begin{figure}[htb]
\begin{center}
\mbox{
\psfig{figure=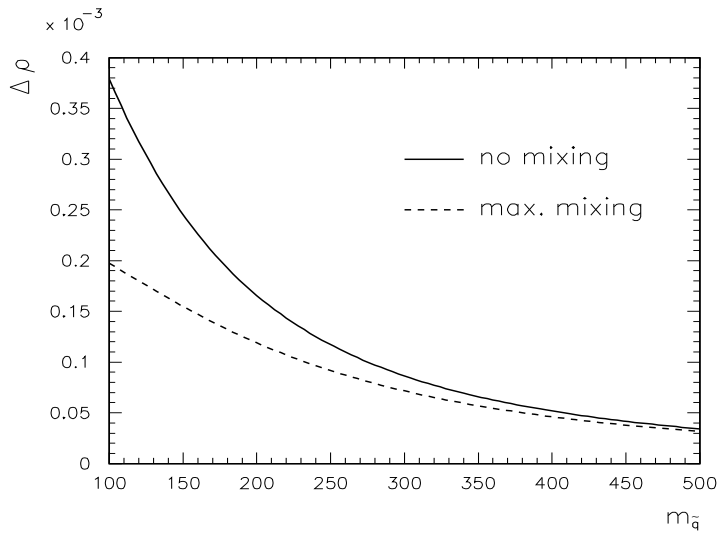,width=9cm,height=6.5cm,bbllx=140pt,bblly=285pt,bburx=450pt,bbury=535pt}}
\end{center}
\caption[]{Gluon exchange contribution to the $\rho$ parameter at two-loop 
order
as a function of $m_{\tilde{q}}$ for the scenarios of \reffi{fig:lett2}.}
\label{fig:lett4}
\end{figure}

{}From the previous expressions, the contribution of the $(\tilde{t}, 
\tilde{b})$ doublet to the $\rho$~parameter, including the two-loop 
gluon exchange and pure scalar quark diagrams, are obtained 
straightforwardly. In the case where the $\tilde{b}$ mixing is neglected, 
the SUSY two-loop contribution is given by an expression similar to 
\refeq{eq:lett5}:
\beqar
\Delta \rho ^{\rm SUSY}_1 &=& \frac{\GF \alpha_s}{4 \sqrt{2} \pi^3} \left[ 
- \sin^2\theta_{\tilde{\Pt}} \cos^2\theta_{\tilde{\Pt}}  
F_1\left( m_{\tilde{\Pt}_1}^2,  m_{\tilde{\Pt}_2}^2 \right) \right. \nonumber \\ 
&&\left. + \cos^2 \theta_{\tilde{\Pt}} F_1 \left( m_{\tilde{\Pt}_1}^2,  
m_{\tilde{\Pb}_L}^2 \right)
+\sin^2 \theta_{\tilde{\Pt}}  F_1 \left( 
m_{\tilde{\Pt}_2}^2,  m_{\tilde{\Pb}_L}^2 \right) \right]. 
\eeqar
The two-loop gluonic SUSY contribution to $\Delta \rho$ is shown in 
\reffi{fig:lett4}
as a function of the common scalar mass $m_{\tilde{q}}$ for the two 
scenarios discussed previously: $\theta_{\tilde{\Pt}} = 0$ and 
$\theta_{\tilde{\Pt}} \simeq -\pi/4$. As can be seen, 
the two-loop contribution 
is of the order of 10 to 15\% of the one-loop result. Contrary to the SM 
case (and to many QCD corrections to electroweak processes in the SM, see
Ref.~\cite{GK} for a review) where the two-loop correction screens 
the one-loop contribution, $\Delta \rho_1^{\rm SUSY}$ has the same sign 
as $\Delta \rho_0^{\rm SUSY}$. For instance, in the case of degenerate 
stops with masses $m_{\tilde{\Pt}} \gg m_{\tilde{\Pb}}$, the 
result is the same as the QCD correction to the $(t,b)$ contribution 
in the SM, but with opposite sign. The gluonic correction to the 
contribution of scalar quarks to the $\rho$ parameter will therefore 
enhance the sensitivity in the search of the virtual effects of scalar
quarks in high-precision electroweak measurements. 

\begin{figure}[htb]
\begin{center}
\mbox{
\psfig{figure=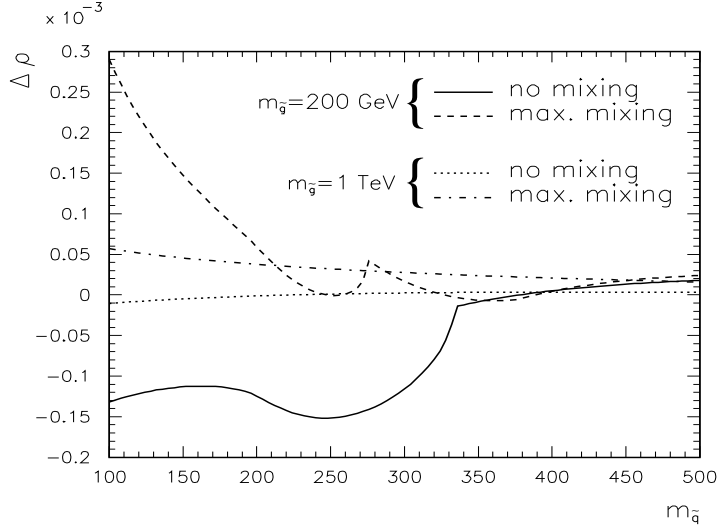,width=9cm,height=6.5cm,bbllx=140pt,bblly=285pt,bburx=450pt,bbury=535pt}}
\end{center}
\caption[]{Contribution of the gluino exchange diagrams to $\Delta \rho_1^
{\rm SUSY}$ for two values of $m_{\tilde{\mathrm g}}$ in the scenarios of 
\reffi{fig:lett2}.}
\label{fig:lett5}
\end{figure} 

The analytical expressions of the contribution of the two-loop diagrams 
with gluino exchange, \reffi{fig:lett3}b, 
to the electroweak gauge-boson self-energies
are very complicated even at zero momentum-transfer. Besides the
fact that the squark mixing leads to a large number of 
contributing diagrams, this is mainly due to the presence of up 
to five particles with different masses in the loops. The lengthy expressions 
will be given elsewhere~\cite{R8}.  It turned out that in general 
the gluino exchange diagrams give smaller contributions compared to gluon 
exchange.  Only for gluino and squark masses close to the 
experimental lower bounds they compete with the gluon exchange 
contributions. In this case, the gluon and gluino contributions add
up to $\sim 30\%$ of the one-loop value for maximal mixing 
(see \reffi{fig:lett5}).
For larger values of $m_{\tilde{\mathrm g}}$, the contribution
decreases rapidly since the gluinos decouple for high masses. 
For vanishing gluino mass, in the limit of exact SUSY, the gluino
exchange contribution reads
$ - \Delta \rho_0^{\rm SM}
\frac{8}{3} \frac{\alpha_s}{\pi}$,
while as mentioned above in the SUSY limit the gluon exchange
contribution of the scalar quarks cancels the one of the quarks.

\section{Conclusions}

In this article higher-order contributions to the relation between 
the vector-boson masses in the SM and the MSSM have been discussed.
In the SM, the Higgs-mass dependence of the two-loop contribution to
$\De r$ has been analyzed. Exact results have been given for the 
$\MH$-dependent corrections associated with the fermions, 
i.e.\ no expansion in $\Mt$, $\MH$ and the gauge-boson masses has 
been made. 
The size of the 
contribution associated with the top quark was found to be about
$10\%$ of the one-loop result and roughly the same as of the higher-order
contributions proportional to $\De\al$, which enter with opposite sign.
These results have been
compared with the result of an expansion up to next-to-leading order in
$\Mt$. Agreement
within about $30 \%$ of the two-loop top-quark correction has been
found, which corresponds to a difference in $\MW$ of
about $4\, \MeV$ in the range $65\, \GeV \leq \MH \leq 1\, \TeV$ of the
Higgs-boson mass.
The Higgs-dependence of the light-fermion contributions leads to a
shift of $\MW$ of up to $4\, \MeV$. The remaining Higgs-dependent
corrections are purely bosonic and have been estimated to give a
relatively small contribution of up to about $2\, \MeV$ in the W-boson
mass.

In the MSSM, the two-loop ${\cal O}(\alpha_s)$ corrections to the
squark-loop contributions to the weak gauge-boson self-energies at zero
momentum-transfer have been calculated and the QCD correction to the
$\rho$~parameter has been derived. The gluonic corrections are of ${\cal
O}(10\%)$: they are positive and increase the sensitivity in the search
for scalar quarks through their virtual effects in high-precision
electroweak observables. The gluino contributions are in general smaller
except for relatively light gluinos and scalar quarks; 
the contribution vanishes for large gluino masses.

\section*{Acknowledgements}

The author thanks M.~Je\.{z}abek and the other organizers of the 
Epiphany Conference for the invitation, the perfect organization and 
their kind hospitality during the Conference. I am grateful to 
my collaborators S.~Bauberger, A.~Djouadi, P.~Gambino, 
S.~Heinemeyer, W.~Hollik and C.~J\"unger, with whom the results
presented here have been worked out.

\end{document}

%% file: DELTAR2.tex
\setlength{\unitlength}{0.1bp}
\special{!
/gnudict 40 dict def
gnudict begin
/Color false def
/Solid false def
/gnulinewidth 5.000 def
/vshift -33 def
/dl {10 mul} def
/hpt 31.5 def
/vpt 31.5 def
/M {moveto} bind def
/L {lineto} bind def
/R {rmoveto} bind def
/V {rlineto} bind def
/vpt2 vpt 2 mul def
/hpt2 hpt 2 mul def
/Lshow { currentpoint stroke M
  0 vshift R show } def
/Rshow { currentpoint stroke M
  dup stringwidth pop neg vshift R show } def
/Cshow { currentpoint stroke M
  dup stringwidth pop -2 div vshift R show } def
/DL { Color {setrgbcolor Solid {pop []} if 0 setdash }
 {pop pop pop Solid {pop []} if 0 setdash} ifelse } def
/BL { stroke gnulinewidth 2 mul setlinewidth } def
/AL { stroke gnulinewidth 2 div setlinewidth } def
/PL { stroke gnulinewidth setlinewidth } def
/LTb { BL [] 0 0 0 DL } def
/LTa { AL [1 dl 2 dl] 0 setdash 0 0 0 setrgbcolor } def
/LT0 { PL [] 0 1 0 DL } def
/LT1 { PL [4 dl 2 dl] 0 0 1 DL } def
/LT2 { PL [2 dl 3 dl] 1 0 0 DL } def
/LT3 { PL [1 dl 1.5 dl] 1 0 1 DL } def
/LT4 { PL [5 dl 2 dl 1 dl 2 dl] 0 1 1 DL } def
/LT5 { PL [4 dl 3 dl 1 dl 3 dl] 1 1 0 DL } def
/LT6 { PL [2 dl 2 dl 2 dl 4 dl] 0 0 0 DL } def
/LT7 { PL [2 dl 2 dl 2 dl 2 dl 2 dl 4 dl] 1 0.3 0 DL } def
/LT8 { PL [2 dl 2 dl 2 dl 2 dl 2 dl 2 dl 2 dl 4 dl] 0.5 0.5 0.5 DL } def
/P { stroke [] 0 setdash
  currentlinewidth 2 div sub M
  0 currentlinewidth V stroke } def
/D { stroke [] 0 setdash 2 copy vpt add M
  hpt neg vpt neg V hpt vpt neg V
  hpt vpt V hpt neg vpt V closepath stroke
  P } def
/A { stroke [] 0 setdash vpt sub M 0 vpt2 V
  currentpoint stroke M
  hpt neg vpt neg R hpt2 0 V stroke
  } def
/B { stroke [] 0 setdash 2 copy exch hpt sub exch vpt add M
  0 vpt2 neg V hpt2 0 V 0 vpt2 V
  hpt2 neg 0 V closepath stroke
  P } def
/C { stroke [] 0 setdash exch hpt sub exch vpt add M
  hpt2 vpt2 neg V currentpoint stroke M
  hpt2 neg 0 R hpt2 vpt2 V stroke } def
/T { stroke [] 0 setdash 2 copy vpt 1.12 mul add M
  hpt neg vpt -1.62 mul V
  hpt 2 mul 0 V
  hpt neg vpt 1.62 mul V closepath stroke
  P  } def
/S { 2 copy A C} def
end
}
\begin{picture}(3600,2160)(0,0)
\special{"
gnudict begin
gsave
50 50 translate
0.100 0.100 scale
0 setgray
/Helvetica findfont 100 scalefont setfont
newpath
-500.000000 -500.000000 translate
LTa
LTb
600 2109 M
63 0 V
2754 0 R
-63 0 V
600 1799 M
63 0 V
2754 0 R
-63 0 V
600 1490 M
63 0 V
2754 0 R
-63 0 V
600 1180 M
63 0 V
2754 0 R
-63 0 V
600 870 M
63 0 V
2754 0 R
-63 0 V
600 561 M
63 0 V
2754 0 R
-63 0 V
600 251 M
63 0 V
2754 0 R
-63 0 V
705 251 M
0 63 V
0 1795 R
0 -63 V
1007 251 M
0 63 V
0 1795 R
0 -63 V
1308 251 M
0 63 V
0 1795 R
0 -63 V
1609 251 M
0 63 V
0 1795 R
0 -63 V
1911 251 M
0 63 V
0 1795 R
0 -63 V
2212 251 M
0 63 V
0 1795 R
0 -63 V
2513 251 M
0 63 V
0 1795 R
0 -63 V
2814 251 M
0 63 V
0 1795 R
0 -63 V
3116 251 M
0 63 V
0 1795 R
0 -63 V
3417 251 M
0 63 V
0 1795 R
0 -63 V
600 251 M
2817 0 V
0 1858 V
-2817 0 V
600 251 L
LT0
1669 1025 M
180 0 V
600 2109 M
8 -13 V
30 -49 V
30 -45 V
30 -40 V
30 -37 V
30 -35 V
30 -32 V
30 -30 V
31 -29 V
30 -26 V
30 -26 V
30 -24 V
30 -23 V
30 -22 V
30 -21 V
30 -20 V
31 -20 V
30 -19 V
30 -18 V
30 -17 V
30 -17 V
30 -17 V
30 -16 V
30 -15 V
31 -15 V
30 -14 V
30 -13 V
30 -13 V
30 -12 V
30 -9 V
30 -10 V
31 -11 V
30 -12 V
30 -12 V
30 -13 V
30 -13 V
30 -14 V
30 -14 V
30 -14 V
31 -14 V
30 -15 V
30 -14 V
30 -15 V
30 -15 V
30 -15 V
30 -15 V
30 -15 V
31 -15 V
30 -16 V
30 -15 V
30 -15 V
30 -15 V
30 -15 V
30 -15 V
30 -16 V
31 -15 V
30 -15 V
30 -15 V
30 -15 V
30 -15 V
30 -15 V
30 -15 V
30 -15 V
31 -15 V
30 -15 V
30 -15 V
30 -15 V
30 -14 V
30 -15 V
30 -15 V
31 -14 V
30 -15 V
30 -14 V
30 -15 V
30 -14 V
30 -15 V
30 -14 V
30 -15 V
31 -14 V
30 -14 V
30 -14 V
30 -15 V
30 -14 V
30 -14 V
30 -14 V
30 -14 V
31 -14 V
30 -14 V
30 -14 V
30 -14 V
30 -13 V
30 -14 V
30 -14 V
30 -14 V
LT1
1669 875 M
180 0 V
600 2109 M
8 -9 V
30 -36 V
30 -33 V
30 -30 V
30 -28 V
30 -26 V
30 -25 V
30 -23 V
31 -22 V
30 -21 V
30 -20 V
30 -19 V
30 -18 V
30 -17 V
30 -16 V
30 -16 V
31 -15 V
30 -14 V
30 -12 V
30 -11 V
30 -12 V
30 -12 V
30 -12 V
30 -13 V
31 -12 V
30 -13 V
30 -13 V
30 -13 V
30 -13 V
30 -13 V
30 -13 V
31 -13 V
30 -12 V
30 -13 V
30 -13 V
30 -12 V
30 -12 V
30 -13 V
30 -12 V
31 -12 V
30 -11 V
30 -12 V
30 -12 V
30 -11 V
30 -12 V
30 -11 V
30 -11 V
31 -11 V
30 -11 V
30 -11 V
30 -11 V
30 -11 V
30 -10 V
30 -11 V
30 -10 V
31 -11 V
30 -10 V
30 -10 V
30 -10 V
30 -10 V
30 -10 V
30 -10 V
30 -9 V
31 -10 V
30 -10 V
30 -9 V
30 -10 V
30 -9 V
30 -9 V
30 -10 V
31 -9 V
30 -9 V
30 -9 V
30 -9 V
30 -9 V
30 -8 V
30 -9 V
30 -9 V
31 -9 V
30 -8 V
30 -9 V
30 -8 V
30 -9 V
30 -8 V
30 -8 V
30 -9 V
31 -8 V
30 -8 V
30 -8 V
30 -8 V
30 -8 V
30 -8 V
30 -8 V
30 -8 V
LT2
1669 725 M
180 0 V
600 2109 M
8 -12 V
30 -44 V
30 -41 V
30 -37 V
30 -35 V
30 -32 V
30 -31 V
30 -28 V
31 -27 V
30 -26 V
30 -24 V
30 -24 V
30 -22 V
30 -21 V
30 -20 V
30 -20 V
31 -19 V
30 -18 V
30 -17 V
30 -17 V
30 -16 V
30 -15 V
30 -14 V
30 -13 V
31 -12 V
30 -12 V
30 -12 V
30 -13 V
30 -13 V
30 -14 V
30 -14 V
31 -14 V
30 -14 V
30 -15 V
30 -14 V
30 -15 V
30 -14 V
30 -15 V
30 -14 V
31 -15 V
30 -14 V
30 -15 V
30 -14 V
30 -14 V
30 -15 V
30 -14 V
30 -14 V
31 -14 V
30 -14 V
30 -14 V
30 -13 V
30 -14 V
30 -14 V
30 -13 V
30 -14 V
31 -13 V
30 -14 V
30 -13 V
30 -13 V
30 -13 V
30 -13 V
30 -13 V
30 -13 V
31 -12 V
30 -13 V
30 -13 V
30 -12 V
30 -13 V
30 -12 V
30 -13 V
31 -12 V
30 -12 V
30 -12 V
30 -12 V
30 -12 V
30 -12 V
30 -12 V
30 -12 V
31 -12 V
30 -11 V
30 -12 V
30 -12 V
30 -11 V
30 -12 V
30 -11 V
30 -11 V
31 -12 V
30 -11 V
30 -11 V
30 -11 V
30 -11 V
30 -12 V
30 -11 V
30 -10 V
LT3
1669 575 M
180 0 V
600 2109 M
8 -13 V
30 -48 V
30 -42 V
30 -39 V
30 -34 V
30 -32 V
30 -28 V
30 -27 V
31 -25 V
30 -23 V
30 -22 V
30 -20 V
30 -19 V
30 -19 V
30 -17 V
30 -17 V
31 -16 V
30 -16 V
30 -15 V
30 -14 V
30 -14 V
30 -14 V
30 -13 V
30 -13 V
31 -12 V
30 -12 V
30 -12 V
30 -11 V
30 -11 V
30 -11 V
30 -10 V
31 -9 V
30 -9 V
30 -8 V
30 -6 V
30 -6 V
30 -8 V
30 -8 V
30 -10 V
31 -9 V
30 -11 V
30 -11 V
30 -12 V
30 -12 V
30 -12 V
30 -13 V
30 -13 V
31 -13 V
30 -14 V
30 -14 V
30 -14 V
30 -14 V
30 -14 V
30 -15 V
30 -14 V
31 -15 V
30 -15 V
30 -15 V
30 -15 V
30 -15 V
30 -15 V
30 -15 V
30 -16 V
31 -15 V
30 -15 V
30 -16 V
30 -15 V
30 -16 V
30 -16 V
30 -15 V
31 -16 V
30 -15 V
30 -16 V
30 -16 V
30 -15 V
30 -16 V
30 -16 V
30 -16 V
31 -15 V
30 -16 V
30 -16 V
30 -15 V
30 -16 V
30 -16 V
30 -16 V
30 -15 V
31 -16 V
30 -16 V
30 -16 V
30 -16 V
30 -15 V
30 -16 V
30 -16 V
30 -15 V
LT4
1669 425 M
180 0 V
600 2109 M
8 -11 V
30 -39 V
30 -34 V
30 -29 V
30 -25 V
30 -21 V
30 -19 V
30 -17 V
31 -15 V
30 -13 V
30 -12 V
30 -11 V
30 -10 V
30 -9 V
30 -8 V
30 -8 V
31 -7 V
30 -7 V
30 -7 V
30 -6 V
30 -6 V
30 -6 V
30 -5 V
30 -6 V
31 -5 V
30 -5 V
30 -5 V
30 -5 V
30 -5 V
30 -5 V
30 -4 V
31 -5 V
30 -4 V
30 -5 V
30 -4 V
30 -3 V
30 -4 V
30 -3 V
30 -1 V
31 0 V
30 0 V
30 -2 V
30 -3 V
30 -4 V
30 -5 V
30 -6 V
30 -6 V
31 -8 V
30 -7 V
30 -9 V
30 -9 V
30 -9 V
30 -10 V
30 -11 V
30 -10 V
31 -11 V
30 -12 V
30 -12 V
30 -12 V
30 -12 V
30 -12 V
30 -13 V
30 -13 V
31 -14 V
30 -13 V
30 -14 V
30 -14 V
30 -14 V
30 -14 V
30 -14 V
31 -14 V
30 -15 V
30 -15 V
30 -15 V
30 -15 V
30 -15 V
30 -15 V
30 -15 V
31 -15 V
30 -16 V
30 -15 V
30 -16 V
30 -16 V
30 -16 V
30 -16 V
30 -16 V
31 -16 V
30 -16 V
30 -16 V
30 -16 V
30 -16 V
30 -17 V
30 -16 V
30 -17 V
stroke
grestore
end
showpage
}
\put(1609,425){\makebox(0,0)[r]{$\Mt = 225 \GeV$}}
\put(1609,575){\makebox(0,0)[r]{$\Mt = 200 \GeV$}}
\put(1609,725){\makebox(0,0)[r]{$\Mt = 150 \GeV$}}
\put(1609,875){\makebox(0,0)[r]{$\Mt = 125 \GeV$}}
\put(1609,1025){\makebox(0,0)[r]{$\Mt = 175 \GeV$}}
\put(3688,151){\makebox(0,0){$\frac{\MH}{\GeV}$}}
\put(40,1080){%
\makebox(0,0)[b]{\shortstack{$\Delta r_{(2),\rm subtr}^{\rm top}$}}%
}
\put(3417,151){\makebox(0,0){1000}}
\put(3116,151){\makebox(0,0){900}}
\put(2814,151){\makebox(0,0){800}}
\put(2513,151){\makebox(0,0){700}}
\put(2212,151){\makebox(0,0){600}}
\put(1911,151){\makebox(0,0){500}}
\put(1609,151){\makebox(0,0){400}}
\put(1308,151){\makebox(0,0){300}}
\put(1007,151){\makebox(0,0){200}}
\put(705,151){\makebox(0,0){100}}
\put(540,251){\makebox(0,0)[r]{$-0.0012$}}
\put(540,561){\makebox(0,0)[r]{$-0.0010$}}
\put(540,870){\makebox(0,0)[r]{$-0.0008$}}
\put(540,1180){\makebox(0,0)[r]{$-0.0006$}}
\put(540,1490){\makebox(0,0)[r]{$-0.0004$}}
\put(540,1799){\makebox(0,0)[r]{$-0.0002$}}
\put(540,2109){\makebox(0,0)[r]{$0$}}
\end{picture}

%% file: DELTARMHDEP.tex
\setlength{\unitlength}{0.1bp}
\special{!
/gnudict 40 dict def
gnudict begin
/Color false def
/Solid false def
/gnulinewidth 5.000 def
/vshift -33 def
/dl {10 mul} def
/hpt 31.5 def
/vpt 31.5 def
/M {moveto} bind def
/L {lineto} bind def
/R {rmoveto} bind def
/V {rlineto} bind def
/vpt2 vpt 2 mul def
/hpt2 hpt 2 mul def
/Lshow { currentpoint stroke M
  0 vshift R show } def
/Rshow { currentpoint stroke M
  dup stringwidth pop neg vshift R show } def
/Cshow { currentpoint stroke M
  dup stringwidth pop -2 div vshift R show } def
/DL { Color {setrgbcolor Solid {pop []} if 0 setdash }
 {pop pop pop Solid {pop []} if 0 setdash} ifelse } def
/BL { stroke gnulinewidth 2 mul setlinewidth } def
/AL { stroke gnulinewidth 2 div setlinewidth } def
/PL { stroke gnulinewidth setlinewidth } def
/LTb { BL [] 0 0 0 DL } def
/LTa { AL [1 dl 2 dl] 0 setdash 0 0 0 setrgbcolor } def
/LT0 { PL [] 0 1 0 DL } def
/LT1 { PL [4 dl 2 dl] 0 0 1 DL } def
/LT2 { PL [2 dl 3 dl] 1 0 0 DL } def
/LT3 { PL [1 dl 1.5 dl] 1 0 1 DL } def
/LT4 { PL [5 dl 2 dl 1 dl 2 dl] 0 1 1 DL } def
/LT5 { PL [4 dl 3 dl 1 dl 3 dl] 1 1 0 DL } def
/LT6 { PL [2 dl 2 dl 2 dl 4 dl] 0 0 0 DL } def
/LT7 { PL [2 dl 2 dl 2 dl 2 dl 2 dl 4 dl] 1 0.3 0 DL } def
/LT8 { PL [2 dl 2 dl 2 dl 2 dl 2 dl 2 dl 2 dl 4 dl] 0.5 0.5 0.5 DL } def
/P { stroke [] 0 setdash
  currentlinewidth 2 div sub M
  0 currentlinewidth V stroke } def
/D { stroke [] 0 setdash 2 copy vpt add M
  hpt neg vpt neg V hpt vpt neg V
  hpt vpt V hpt neg vpt V closepath stroke
  P } def
/A { stroke [] 0 setdash vpt sub M 0 vpt2 V
  currentpoint stroke M
  hpt neg vpt neg R hpt2 0 V stroke
  } def
/B { stroke [] 0 setdash 2 copy exch hpt sub exch vpt add M
  0 vpt2 neg V hpt2 0 V 0 vpt2 V
  hpt2 neg 0 V closepath stroke
  P } def
/C { stroke [] 0 setdash exch hpt sub exch vpt add M
  hpt2 vpt2 neg V currentpoint stroke M
  hpt2 neg 0 R hpt2 vpt2 V stroke } def
/T { stroke [] 0 setdash 2 copy vpt 1.12 mul add M
  hpt neg vpt -1.62 mul V
  hpt 2 mul 0 V
  hpt neg vpt 1.62 mul V closepath stroke
  P  } def
/S { 2 copy A C} def
end
}
\begin{picture}(3600,2160)(0,0)
\special{"
gnudict begin
gsave
50 50 translate
0.100 0.100 scale
0 setgray
/Helvetica findfont 100 scalefont setfont
newpath
-500.000000 -500.000000 translate
LTa
600 251 M
2817 0 V
LTb
600 251 M
63 0 V
2754 0 R
-63 0 V
600 516 M
63 0 V
2754 0 R
-63 0 V
600 782 M
63 0 V
2754 0 R
-63 0 V
600 1047 M
63 0 V
2754 0 R
-63 0 V
600 1313 M
63 0 V
2754 0 R
-63 0 V
600 1578 M
63 0 V
2754 0 R
-63 0 V
600 1844 M
63 0 V
2754 0 R
-63 0 V
600 2109 M
63 0 V
2754 0 R
-63 0 V
705 251 M
0 63 V
0 1795 R
0 -63 V
1007 251 M
0 63 V
0 1795 R
0 -63 V
1308 251 M
0 63 V
0 1795 R
0 -63 V
1609 251 M
0 63 V
0 1795 R
0 -63 V
1911 251 M
0 63 V
0 1795 R
0 -63 V
2212 251 M
0 63 V
0 1795 R
0 -63 V
2513 251 M
0 63 V
0 1795 R
0 -63 V
2814 251 M
0 63 V
0 1795 R
0 -63 V
3116 251 M
0 63 V
0 1795 R
0 -63 V
3417 251 M
0 63 V
0 1795 R
0 -63 V
600 251 M
2817 0 V
0 1858 V
-2817 0 V
600 251 L
LT0
3025 1180 M
180 0 V
600 251 M
8 15 V
30 59 V
30 54 V
30 49 V
30 46 V
30 43 V
30 41 V
30 38 V
31 37 V
30 34 V
30 33 V
30 32 V
30 30 V
30 29 V
30 27 V
30 27 V
31 26 V
30 25 V
30 24 V
30 23 V
30 23 V
30 21 V
30 22 V
30 20 V
31 20 V
30 20 V
30 19 V
30 18 V
30 18 V
30 17 V
30 18 V
31 16 V
30 17 V
30 16 V
30 15 V
30 16 V
30 15 V
30 14 V
30 15 V
31 14 V
30 14 V
30 13 V
30 13 V
30 13 V
30 13 V
30 13 V
30 12 V
31 12 V
30 12 V
30 12 V
30 12 V
30 11 V
30 11 V
30 11 V
30 11 V
31 11 V
30 10 V
30 11 V
30 10 V
30 10 V
30 10 V
30 10 V
30 9 V
31 10 V
30 9 V
30 10 V
30 9 V
30 9 V
30 9 V
30 9 V
31 8 V
30 9 V
30 8 V
30 9 V
30 8 V
30 8 V
30 8 V
30 8 V
31 8 V
30 8 V
30 8 V
30 8 V
30 7 V
30 8 V
30 7 V
30 7 V
31 8 V
30 7 V
30 7 V
30 7 V
30 7 V
30 7 V
30 7 V
30 7 V
LT1
3025 980 M
180 0 V
600 251 M
8 14 V
30 55 V
30 49 V
30 46 V
30 43 V
30 41 V
30 38 V
30 35 V
31 34 V
30 32 V
30 31 V
30 30 V
30 28 V
30 27 V
30 26 V
30 25 V
31 24 V
30 23 V
30 22 V
30 22 V
30 21 V
30 21 V
30 20 V
30 19 V
31 19 V
30 18 V
30 18 V
30 17 V
30 17 V
30 17 V
30 16 V
31 16 V
30 15 V
30 15 V
30 15 V
30 14 V
30 14 V
30 13 V
30 13 V
31 13 V
30 13 V
30 12 V
30 12 V
30 12 V
30 11 V
30 12 V
30 11 V
31 11 V
30 10 V
30 11 V
30 10 V
30 10 V
30 10 V
30 10 V
30 9 V
31 10 V
30 9 V
30 9 V
30 9 V
30 9 V
30 8 V
30 9 V
30 8 V
31 8 V
30 9 V
30 8 V
30 7 V
30 8 V
30 8 V
30 7 V
31 8 V
30 7 V
30 7 V
30 8 V
30 7 V
30 7 V
30 6 V
30 7 V
31 7 V
30 7 V
30 6 V
30 7 V
30 6 V
30 6 V
30 6 V
30 7 V
31 6 V
30 6 V
30 5 V
30 6 V
30 6 V
30 6 V
30 5 V
30 6 V
LT2
3025 780 M
180 0 V
600 251 M
8 16 V
30 60 V
30 56 V
30 51 V
30 48 V
30 44 V
30 43 V
30 39 V
31 38 V
30 36 V
30 34 V
30 33 V
30 31 V
30 30 V
30 29 V
30 28 V
31 27 V
30 26 V
30 25 V
30 24 V
30 23 V
30 23 V
30 22 V
30 21 V
31 21 V
30 21 V
30 19 V
30 20 V
30 19 V
30 18 V
30 18 V
31 18 V
30 17 V
30 17 V
30 16 V
30 16 V
30 15 V
30 15 V
30 15 V
31 14 V
30 14 V
30 14 V
30 13 V
30 14 V
30 13 V
30 12 V
30 13 V
31 12 V
30 12 V
30 11 V
30 12 V
30 11 V
30 11 V
30 11 V
30 11 V
31 10 V
30 11 V
30 10 V
30 10 V
30 10 V
30 10 V
30 9 V
30 10 V
31 9 V
30 9 V
30 9 V
30 9 V
30 9 V
30 8 V
30 9 V
31 8 V
30 8 V
30 9 V
30 8 V
30 8 V
30 7 V
30 8 V
30 8 V
31 7 V
30 8 V
30 7 V
30 7 V
30 8 V
30 7 V
30 7 V
30 7 V
31 7 V
30 6 V
30 7 V
30 7 V
30 6 V
30 7 V
30 6 V
30 6 V
LT3
3025 580 M
180 0 V
600 251 M
8 16 V
30 61 V
30 57 V
30 52 V
30 48 V
30 46 V
30 43 V
30 40 V
31 38 V
30 37 V
30 35 V
30 33 V
30 32 V
30 30 V
30 30 V
30 28 V
31 27 V
30 27 V
30 25 V
30 25 V
30 24 V
30 23 V
30 22 V
30 22 V
31 22 V
30 20 V
30 21 V
30 19 V
30 20 V
30 19 V
30 18 V
31 18 V
30 17 V
30 17 V
30 17 V
30 16 V
30 16 V
30 15 V
30 15 V
31 15 V
30 14 V
30 15 V
30 13 V
30 14 V
30 13 V
30 13 V
30 13 V
31 12 V
30 12 V
30 12 V
30 12 V
30 12 V
30 11 V
30 11 V
30 11 V
31 11 V
30 11 V
30 10 V
30 11 V
30 10 V
30 10 V
30 10 V
30 9 V
31 10 V
30 9 V
30 9 V
30 10 V
30 9 V
30 8 V
30 9 V
31 9 V
30 8 V
30 9 V
30 8 V
30 8 V
30 8 V
30 8 V
30 8 V
31 8 V
30 8 V
30 7 V
30 8 V
30 7 V
30 7 V
30 8 V
30 7 V
31 7 V
30 7 V
30 7 V
30 6 V
30 7 V
30 7 V
30 6 V
30 7 V
stroke
grestore
end
showpage
}
\put(2965,580){\makebox(0,0)[r]{$\Delta r_{\rm subtr}$}}
\put(2965,780){\makebox(0,0)[r]{$\Delta r_{(1),\rm subtr}+     \Delta r_{(2),\rm subtr}^{{\rm top}}+       \Delta r_{(2),\rm subtr}^{\Delta \alpha }$}}
\put(2965,980){\makebox(0,0)[r]{$\Delta r_{(1),\rm subtr}+ \Delta r_{(2),\rm subtr}^{\rm top}$}}
\put(2965,1180){\makebox(0,0)[r]{$\Delta r_{(1),\rm subtr}$}}
\put(3688,151){\makebox(0,0){$\frac{\MH}{\GeV}$}}
\put(40,1180){%
\makebox(0,0)[b]{\shortstack{$\Delta r_{\rm subtr}$}}%
}
\put(3417,151){\makebox(0,0){1000}}
\put(3116,151){\makebox(0,0){900}}
\put(2814,151){\makebox(0,0){800}}
\put(2513,151){\makebox(0,0){700}}
\put(2212,151){\makebox(0,0){600}}
\put(1911,151){\makebox(0,0){500}}
\put(1609,151){\makebox(0,0){400}}
\put(1308,151){\makebox(0,0){300}}
\put(1007,151){\makebox(0,0){200}}
\put(705,151){\makebox(0,0){100}}
\put(540,2109){\makebox(0,0)[r]{$0.014$}}
\put(540,1844){\makebox(0,0)[r]{$0.012$}}
\put(540,1578){\makebox(0,0)[r]{$0.010$}}
\put(540,1313){\makebox(0,0)[r]{$0.008$}}
\put(540,1047){\makebox(0,0)[r]{$0.006$}}
\put(540,782){\makebox(0,0)[r]{$0.004$}}
\put(540,516){\makebox(0,0)[r]{$0.002$}}
\put(540,251){\makebox(0,0)[r]{$0$}}
\end{picture}